# Silicon substituted hydroxyapatite/VEGF scaffolds stimulate bone regeneration in osteoporotic sheep

L. Casarrubios[1], N. Gómez-Cerezo[2,3], S. Sánchez-Salcedo[2,3], M.J. Feito[1], M.C. Serrano[4], M. Saiz-Pardo[5], L. Ortega[5], D. de Pablo[5], I. Díaz-Güemes[6], B. Fernández-Tomé[6], S. Enciso[6], F. M. Sánchez-Margallo[6], M.T. Portolés[1*], D. Arcos[2,3*], M. Vallet-Regí[2,3*]

1. Departamento de Bioquímica y Biología Molecular, Facultad de Ciencias Químicas, Universidad Complutense de Madrid, Instituto de Investigación Sanitaria del Hospital Clínico San Carlos (IdISSC), Ciudad Universitaria, 28040 Madrid, Spain.

2. Departamento de Química en Ciencias Farmacéuticas, Facultad de Farmacia, Universidad Complutense de Madrid, Instituto de Investigación Sanitaria del Hospital 12 de Octubre i+12, Plaza Ramón y Cajal s/n, 28040 Madrid, Spain

3. CIBER de Bioingeniería Biomateriales y Nanomedicina (CIBER-BBN), Spain.

4. Instituto de Ciencia de Materiales de Madrid (ICMM), Consejo Superior de Investigaciones Científicas (CSIC), 28049 Madrid, Spain

5. Servicio de Anatomía Patológica, Hospital Clínico San Carlos, Facultad de Medicina, Universidad Complutense de Madrid, Instituto de Investigación Sanitaria del Hospital Clínico San Carlos (IdISSC), 28040 Madrid, Spain

6. Centro de Cirugía de Mínima Invasión Jesús Usón, Cáceres, Spain

* Corresponding authors: portoles@quim.ucm.es (M.T. Portolés), arcosd@ucm.es (D. Arcos) and vallet@ucm.es (M. Vallet-Regí)

**Abstract**

Silicon-substituted hydroxyapatite (SiHA) macroporous scaffolds have been prepared by robocasting. In order to optimize their bone regeneration properties, we have manufactured these scaffolds presenting different microstructures: nanocrystalline and crystalline. Moreover, their surfaces have been decorated with vascular endothelial growth factor (VEGF) to evaluate the potential coupling between vascularization and bone regeneration. *In vitro* cell culture tests evidence that nanocrystalline SiHA hinders pre-osteblast proliferation, whereas the presence of VEGF enhances the biological functions of both endothelial cells and pre-osteoblasts. The bone regeneration capability has been evaluated using an osteoporotic sheep model. *In vivo* observations strongly correlate with *in vitro* cell culture tests. Those scaffolds made of nanocrystalline SiHA were colonized by fibrous tissue, promoted inflammatory response and fostered osteoclast recruitment. These observations discard nanocystalline SiHA as a suitable material for bone regeneration purposes. On the contrary, those scaffolds made of crystalline SiHA and decorated with VEGF exhibited bone regeneration properties, with high ossification degree, thicker trabeculae and higher presence of osteoblasts and blood vessels. Considering these results, macroporous scaffolds made of SiHA and decorated with VEGF are suitable bone grafts for regeneration purposes, even in adverse pathological scenarios such as osteoporosis.

# 1. Introduction

Since Ruys proposed in 1993 the incorporation of silicon within the hydroxyapatite structure [1], silicon substituted hydroxyapatites (SiHA) have been one of the most investigated bioceramics in the field of dental and orthopedic applications. Based on the positive effects of silicon on bone [2] and the osteoregenerative properties of silica based bioactive glasses [3], the interest for the characterization and biological evaluation of these compounds has encouraged many researchers for preparing different silicon substituted calcium phosphates [4-6]. In SiHA, silicon substitutes phosphorous in the hydroxyapatite crystal lattice. This substitution leads to crystallo-chemical disorders in the apatite unit cell [7-10], surface charge modifications [11] and microstructural changes that enhance the solubility and bioactive behavior of these bioceramics [12-14], although this point has caused some controversy [15].

SiHA has been conventionally prepared by a controlled precipitation method and subsequent sintering, which allows for solid state reaction completion and enhancement of mechanical properties [16,17]. However, several authors have addressed the preparation of HA and SiHA using low temperature routes for obtaining bioceramics with small grain sizes or even nanoparticles [18-20]. Grain size reduction increases the bioceramic solubility [21], thus releasing calcium cations and soluble silica species that could enhance osteoblast function. Webster et al. investigated the potential use of nanostructured bioceramics for orthopedic and dental applications and demonstrated that nanophase alumina, titania and hydroxypatite [22,23] enhance osteoblast adhesion compared to the same compounds with larger grain size. Previous *in vitro* results have evidenced that nanocrystalline SiHA (NanoSiHA) delays osteoclast differentiation and resorptive activity [24] and promotes macrophage polarization towards a more reparative M2 phenotype [25]. However, most of these studies have been carried out under *in vitro* conditions and only a few *in vivo* studies involving nanosized SiHAs have been reported [26-28].

Additive manufacturing processes allow the control of the implant morphology and internal architecture of the scaffolds [29]. In this context, SiHAs have been manufactured as macroporous scaffolds through techniques such as stereolithography and robocasting, evidencing that these techniques allow the optimization of biological processes such as vascularization and cell migration and proliferation among others, through the control of the size, volume and connectivity of the pores [30-36].

Finally, decoration of scaffolds with growth factors is an effective strategy for modifying the biological behavior of bone implants. Considering the importance of vascularization during bone regeneration processes [37], the adhesion of vascular endothelial growth factor (VEGF) has become one of the most useful alternatives to improve the scaffolds performance [38-41]. Recently, VEGF has been adsorbed on NanoSiHA disks improving the adhesion and proliferation of endothelial progenitor cells (EPC) [42]. Since VEGF is involved in both angiogenesis and osteoblast maturation, in the present study 3D macroporous NanoSiHA and SiHA scaffolds with high interconnectivity and adsorbed VEGF were prepared and characterized in order to evaluate *in vitro* their ability to support endothelial cell colonization and MC3T3-E1 pre-osteoblast proliferation and differentiation. Then, *in vivo* studies were carried out by implanting these scaffolds into a cavitary defect of osteoporotic sheep and evaluating the new bone and blood vessels formation. The effects of the SiHA crystallite size, i.e. micro or nanometric scales, have been also evaluated. Our hypothesis is that 3D printed macroporous SiHA based scaffolds decorated with VEGF could improve bone regeneration in a scenario of osteoporosis, whereas the microstructure of the bioceramics could significantly modify the *in vivo* behavior of the scaffolds due to their different cell-surface interactions.

## 2. Materials and methods

### 2.1. Synthesis of materials.

SiHAs with nominal formula $Ca_{10}(PO_4)_{5.75}(SiO_4)_{0.25}(OH)_{1.75}\square_{0.25}$ (being $\square$ vacancies at the hydroxyl sites) were prepared by aqueous precipitation reaction of $Ca(NO_3)_2 \cdot 4H_2O$, $(NH_4)_2HPO_4$ and $Si(CH_3CH_2O)_4$ solutions as described elsewhere [43] (see supplementary data for further information). The obtained precipitate was dried and subsequently treated at 700°C under air atmosphere to remove the nitrates without introducing significant changes in the microstructure of the precipitated powder.

The Si-HA grains were suspended in ethanol, milled in a rotatory mill and finally sieved, collecting the powder fraction below 32 μm.

**2.2. Fabrication of NanoSiHA and SiHA scaffolds**

Cylindrical scaffolds with 10 mm in height and 10 mm in diameter were prepared by robocasting. For this purpose, 0.5 g of polyvinyl alcohol (PVA) were dissolved in 2.8 mL of D.I. water at 80º C for 2 h. PVA is a biocompatible and biodegradable polymer, which has been used in different bio-applications as drug release, tissue engineering and enzyme immobilization [44-47]. Subsequently 3 g of SiHA particles were incorporated in this solution. Scaffolds were designed as a computer assisted design (CAD) file and prepared as a lattice of rods stacked with tetragonal symmetry and constructed via polymerization of this mixture using a robotic deposition apparatus 3-D Bioplotter$^{TM}$. The scaffolds were kept at 40º C overnight and subsequently heated at 150º C for 30 min to induce heat-activated cross-linking of PVA, without modifying the microstructure of the ceramic component. These scaffolds have a weight of about 305 mg and are referred as NanoSiHA. In order to obtain highly crystalline SiHA scaffolds, NanoSiHA scaffolds were thermally treated at 1150ºC for 3 h. These specimens have a weight of about 260 mg and are referred as SiHA scaffolds. (see supporting information for complete details).

**2.3. Preparation of VEGF-decorated scaffolds**

NanoSiHA and SiHA scaffolds were introduced into 24 well culture plates (CULTEK S. L. U., Madrid, Spain) and sterilized under ultraviolet light during 1 hour for each side in a sterile environment. Adsorption of VEGF on scaffold surfaces was carried out through non-covalent binding by incubation of each scaffold with 5 μg/mL of VEGF-A (VEGF-121, 583204, Biolegend, San Diego, CA, USA) in phosphate buffered saline (PBS, Sigma-Aldrich, St Louis, MO, USA) at 4 ºC for different times. The samples were stored at 4ºC for 24 h before the *in vitro* and *in vivo* studies. After 0 min, 10 min, 30 min and 24 hours of incubation, the concentration of free VEGF in the supernatants was analyzed by using an enzyme linked immunosorbent assay (ELISA, Cloud-Clone Corp, USA). The adsorbed VEGF amount was indirectly calculated as the difference between the free VEGF levels at the initial time and after each incubation time. Scaffolds decorated with VEGF are referred as NanoSiHA/VEGF and SiHA/VEGF.

**2.4. Characterization techniques**

X-ray diffraction experiments were carried out in a X'pert Philips diffractometer using the KαCu radiation ($\lambda$ =1,5418 Å) with a step size of 0.03 $2\theta°$ and 3 s of counting time. Crystallite size was calculated as an averaged value of several diffraction maxima using the Scherrer formula: Crystallite size (average) = $K\lambda/(B_{struct.} \cos\theta)$, where K is the shape constant (in our case 0.9), $\theta$ the diffraction angle and $B_{struct}$ describes the structural broadening, which is the difference in integral profile width between a standard and the sample to be analyzed. The standard used was a well crystallized sample of $LaB_6$. Fourier Transform Infrared (FTIR) spectroscopy was carried out by using a Nicolet Nexus instrument (Thermo Fisher Scientific) equipped with an attenuated total reflectance device (Thermo Electron Scientific Instruments LLC, Madison, WI USA). The textural properties of the scaffolds were determined by nitrogen adsorption with a Micromeritics ASAP 2020 equipment (Micromeritics Co., Norcross, USA). To perform the $N_2$ adsorption measurements, the samples were previously degassed under vacuum for 24 h, at 45º C. Scanning electron microscopy (SEM) was performed with a JEOL F-6335 microscope, operating at 20 kV.

**2.5. Determination of the in vitro VEGF desorption from NanoSiHA and SiHA**

To measure the VEGF spontaneous desorption from NanoSiHA and SiHA (previously incubated for 24 hours into 24 well culture plates with 5µg/ml of this factor), the samples were incubated with 500 µl of PBS for 24, 60 and 96 hours and the free VEGF in the medium was analyzed by ELISA.

**2.6. Culture of endothelial cells**

Mature endothelial cells $EC_2$, obtained from porcine peripheral blood as previously described [48], were seeded on NanoSiHA and SiHA scaffolds at a density of $3x10^5$ cells/mL in EGM-2 (Lonza, Walkersville, MD, USA). In order to evaluate the effects of the VEGF immobilized on the scaffold surface, EGM-2 without VEGF was used for the culture of $EC_2$ on these scaffolds. All the samples were incubated under 5% $CO_2$ atmosphere at 37 ºC for 7 days.

**2.7. Culture of MC3T3-E1 pre-osteoblasts**

In order to evaluate the proliferation of murine pre-osteoblastic MC3T3-E1 cells (subclone 4, CRL-2593, ATCC, Manassas, VA, USA) on the scaffolds, $3\times10^5$ cells/mL were seeded in alpha-Minimum Essential Medium (alpha-MEM, Sigma Chemical Company, St. Louis, MO, USA) supplemented with 10% fetal bovine serum (FBS, Gibco, BRL), 1 mM L-glutamine (BioWhittaker Europe, Belgium), penicillin (200 µg/mL, BioWhittaker Europe, Belgium), and streptomycin (200 µg/mL, BioWhittaker Europe, Belgium). In order to evaluate the differentiation of MC3T3-E1 pre-osteoblasts on the scaffolds, $3\times10^5$ cells/mL were seeded in differentiation culture medium (alpha-MEM supplemented with 10% FBS, 50 µg/mL β-glycerolphosphate, 10 mM L-ascorbic acid, 1 mM L-glutamine, penicillin and streptomycin). All the samples were incubated under 5% $CO_2$ atmosphere at 37 ºC for 7 days.

**2.8. Cell proliferation assay (CCK-8)**

Proliferation of both endothelial cells ($EC_2$) and MC3T3 pre-osteoblasts was measured using the Cell Counting Kit-8 (CCK-8) protocol (Sigma-Aldrich, St Louis, MO, USA). After incubation for 3-4 hours under 5% $CO_2$ atmosphere at 37 ºC, samples of 100 µL were collected into 96 well culture plates (Nunc Brand, Rochester, NY, USA) and absorbance was measured at 450 nm.

**2.9. Alkaline phosphatase (ALP) activity**

The ALP activity was used as the key differentiation marker in assessing the osteoblast phenotype. ALP activity was measured with a commercial kit (SpinReact S.A., Spain) and normalized to the cell protein content, as determined by the Bradford's method using bovine serum albumin as a standard.

**2.10. Morphological studies by SEM**

Cells attached to the scaffolds were fixed by incubation with 2.5% glutaraldehyde (Merck KGaA, Darmstadt, Germany) for 1 h at 4 ºC. Then, successive dehydration steps were carried out by slow water replacement, using a series of ethanol solutions (30%, 50%, 70%, 80% and 90%) for 10 min and a final dehydration in absolute ethanol for 10 min, allowing the samples to dry at room temperature. Afterwards, the pieces were mounted on stubs and coated in vacuum with gold–palladium. Samples were then examined with a JEOL JSM-6400 SEM (Centro Nacional de Microscopía Electrónica, Madrid, Spain).

**2.11. *In vivo* studies in an osteoporotic sheep model**

This study was approved by our Institutional Ethical Committee following the guidelines of the current normative (Directive 2010/63/EU of the European Parliament and the Council of September 22, 2010, on the protection of animals used for scientific purposes). To reproduce conditions similar to osteoporosis in humans, laparoscopic bilateral ovariectomy was performed in six 4-year-old female Merino sheep following the same protocol described elsewhere [28]. A low-calcium diet and corticosteroids were administered until the end of the study (see supporting information for further details).

Six months after the ovariectomy, the biomaterials were blindly implanted in the sheep under aseptic conditions and following the same anesthetic protocol previously described [28]. Six cylindrical defects (10x13 mm) were created in each sheep by drilling the cancellous bone at different limbs locations, under continuous irrigation with cold sterile saline. Two defects in each sheep were left empty as control. Once the biomaterials were randomly implanted, the muscular and subcutaneous tissue was approximated with absorbable monofilament suture and the skin with absorbable braided suture. Immediately after the surgical procedure and before the sample removal, a computed tomography (CT) scan was performed (Figure S5 in supporting information). The sample size for each type of scaffold was n = 6.

**2.12. Histological processing**

After 12 weeks of implantation, the bone segments containing the defect were dissected out and fixed by immersion in 96% ethanol. Bone segments were processed as described elsewhere [49] and histological sections were stained with Hematoxylin-Eosin (Agilent Dako Coverstainer for H&E) (see supplementary data for further information). Images were obtained with a Leica DMD1008 and an Olympus BX40 microscopes and analyzed with ImageJ 1.x to calculate bone ingrowth area, trabeculae thickness and number of blood vessels. For the evaluation of the inflammatory component and osteoblast/osteoclast presence, we used a five graded scale based on the density and distribution (absent/mild/moderate/marked/ very marked, scored 0 to 4). Multinucleated cells sited in a resorption lacuna were identified as osteoclasts and oval cells with abundant blue-grey cytoplasm and perinuclear hofs, rimming the forming bone were identified as osteoblasts. In the case of vascularization, we used a four graded scale based on the density of blood vessels (absent/mild/moderate/marked/, scored 0 to 3) (see Table S1 in supporting information)

**2.13. Statistics**

*In vitro* cell culture data are expressed as means ± standard deviations of a representative of three experiments carried out in triplicate. Statistical analysis was performed using the Statistical Package for the Social Sciences (SPSS) version 19 software. Statistical comparisons for both *in vitro* and *in vivo* data were made by analysis of variance (ANOVA), with the assumption of data normality. Homoscedasticity was evaluated by Levene test. Either Scheffé or Games-Howell tests were used for *post hoc* evaluations of differences between groups depending on the homogeneity of the variance. In all the statistical evaluations, $p < 0.05$ was considered as statistically significant.

**3. Results**

**3.1. Characterization of NanoSiHA and SiHA scaffolds**

XRD patterns of NanoSiHA and SiHA samples showed a unique apatite-like phase (Figure S1 in supporting information). All the diffraction maxima can be assigned to a hexagonal spatial group P63/m. XRD patterns evidence significant differences from a microstructural point of view. NanoSiHA shows diffraction maxima with larger Full Width at Half Maximum (FWHM) compared to SiHA (Figure S1.b in supporting information). This fact indicates that NanoSiHA exhibits smaller average crystallite size. The crystallite sizes calculated by the Scherrer formula from ten different diffraction maxima of each pattern were 30.5 ± 1.2 and 96.2 ± 2.8 nm for NanoSiHA and SiHA, respectively.

FTIR spectra of NanoSiHA and SiHA (Figure S2 in supporting information) show absorption bands corresponding to the bending vibration mode of O-P-O bonds, the stretching mode of P-O and the librational and stretching modes of O-H in HA. In addition, NanoSiHA also shows absorption bands corresponding to the functional groups of PVA and bands corresponding to $CO_3^{2-}$ groups at 1550-1630 $cm^{-1}$ commonly included in apatites thermally treated below the sintering temperature.

Figure 1 shows the SEM micrographs corresponding to NanoSiHA and SiHA scaffolds. Low magnification micrographs show an almost identical macro-architecture in both samples, with orthogonal macroporores of 500 μm. However, higher magnification images evidence significant differences in the microstructure of the samples. NanoSiHA scaffold is formed by very small particles of a few nanometers that keep together by means of a continuous phase of PVA. On the contrary, the microstructure of SiHA scaffolds is described by the presence of larger polyhedral crystals (around 2-3 micrometers in size) and with some porosity at the macropore range. Due to the burning out of PVA during the thermal treatment, SiHA scaffolds showed a loss of weight about 15%, in agreement with the thermogravimetric analysis carried out on NanoSiHA (see Figure S3 in supporting information) and a size reduction of less than 5% in both diameter and height.

Nitrogen adsorption indicates that NanoSiHA scaffolds show surface areas, porosities and pore sizes at the mesopore range (Table 1) in the range of hydroxyapatites treated at low temperatures [50]. However, no porosity at the mesopore level could be observed by for SiHA scaffolds.

### 3.2. Determination of VEGF adsorption and release

VEGF adsorption on NanoSiHA and SiHA scaffolds reached over 94% after 10 minutes in both cases and maintained after 24 hours (Figure S4.a in supporting information). No significant differences were observed in the amount of VEGF adsorbed between NanoSiHA and SiHA scaffolds. Since NanoSiHA and SiHA scaffolds have a weight of 305 mg and 260 mg respectively, the VEGF/scaffold weight ratios were determined as $1,63 \cdot 10^{-5}$ and $1,92 \cdot 10^{-5}$ for NanoSiHA and SiHA, respectively. Desorbed VEGF detected in the medium of NanoSiHA and SiHA during the release test (Figure S4.b) reached levels of less than 2 ng/ml after 96 hours in all the cases. This VEGF desorption meant 0.002 wt % and 0.02 wt % respect to the VEGF adsorbed on NanoSiHA and SiHA, respectively, thus indicating the effective immobilization of VEGF.

### 3.3. *In vitro* response of endothelial cells

Figure 2 shows the proliferation of EC$_2$ endothelial cells on NanoSiHA and SiHA scaffolds without (white) and with (black) adsorbed VEGF after 7 days of culture. The EC$_2$ growth on NanoSiHA/VEGF and SiHA/VEGF scaffolds was significantly higher than on the scaffolds without VEGF (*$p < 0.05$ and **$p < 0.01$, respectively), evidencing the importance of the adsorption of this growth factor in order to improve the endothelial cell proliferation and survival. There were no statistically significant differences between NanoSiHA and SiHA scaffolds in each situation (without or with VEGF). SEM observations in Figure 3 evidence that endothelial cells adhered and proliferated on the surface of all these scaffolds after 7 days of culture. However, better EC$_2$ cell adhesion and growth were observed on NanoSiHA/VEGF and SiHA/VEGF scaffolds compared to NanoSiHA and SiHA, respectively.

### 3.4. *In vitro* response of MC3T3 pre-osteoblasts

The response of MC3T3 pre-osteoblasts cultured on NanoSiHA and SiHA scaffolds without and with adsorbed VEGF was evaluated after 7 days by measuring cell proliferation and the ALP activity. Figure 4 shows that SiHA scaffolds exhibited a significant increase of pre-osteoblasts proliferation ($^{\phi\phi\phi}p < 0.005$) and differentiation ($^{\phi\phi\phi}p < 0.005$) on their surface in comparison with NanoSiHA samples. The presence of VEGF adsorbed on these two types of scaffolds did not improve the pre-osteoblast proliferation but induced a significant increase of pre-osteoblast differentiation on SiHA/VEGF scaffolds (*$p < 0.05$), as it is indicated by the measurement of ALP activity.

SEM studies were also carried out with MC3T3 pre-osteoblasts cultured on NanoSiHA and SiHA scaffolds with and without adsorbed VEGF. Figure 5 evidences that pre-osteoblasts adhered and spread on all these materials after 7 days of culture showing their typical morphology.

### 3.5. CT scanning from implanted osteoporotic sheep

CT scan images after 12 weeks after implantation are shown in Figure 6. The control defect remains unhealed, pointing out that the defect model chosen does not allow for bone self-healing under this osteoporosis scenario. Scaffolds treated at high temperature (SiHA and SiHA/VEGF) keep most of their initial architecture, being well integrated within bone. No scaffold protrusion was observed in any sheep. On the contrary, scaffolds made of nanocrystalline SiHA (NanoSiHA and NanoSiHA/VEGF) appear as highly degraded implants after 12 weeks of implantation, with some particles migrating out of the bone defect area.

**3.6. Histological evaluation and histomorphometric quantification**

Histological images of SiHA scaffolds (Figure 7a) show reparative peripheral bone surrounding the scaffolds and new bone formation within the macropores in several inner regions. A higher magnification image (Figure 7b) shows that the newly formed bone is in intimate contact with SiHA bioceramics as well as the presence of an osteoblast border around the newly formed trabeculae. SiHA/VEGF scaffolds (Figure 7c) evidence a much higher bone growth at both peripheral regions and inner macropores of the implant. In fact, most of the scaffold is filled with new bone tissue, formed by thick trabeculae that are surrounded by osteoblasts (Figure 7 d). Moreover, a high presence of blood vessels is observed in this tissue (see Figure S6 in supporting information). The histological image for NanoSiHA scaffolds (Figure 7e) shows a small amount of reparative mineralized bone at the peripheral site whereas the inner porosity is colonized only by fibrous tissue. A higher magnification image (Figure 7f) shows a significant number of multinucleated cells, which can be identified as osteoclasts, near to the implanted material. In addition, some areas of the scaffolds showed the presence of inflammatory components (Figure 7f, inset). NanoSiHA/VEGF scaffolds showed a very similar behavior to NanoSiHA, although a higher amount of reparative peripheral bone (Figure 7g) and more blood vessels (Figure 7h) could be observed.

The volume of the scaffolds colonized by newly formed bone is shown in Figure 8a. It can be observed that SiHA/VEGF scaffolds induce a significantly higher bone regeneration than SiHA (taken as positive control). On the contrary, NanoSiHA and NanoSiHA/VEGF showed ossification percentages much lower than SiHA. These data closely correlate with the thickness of trabeculae (Figure 8b). The newly formed bone in SiHA/VEGF exhibits significantly thicker trabeculae compared to SiHA, whereas NanoSiHA and NanoSiHA/VEGF show trabeculae significantly thinner than SiHA.

The four series of scaffolds were scored regarding the amount of osteoblasts (Figure 8c), osteoclasts (Figure 8d), inflammatory component (Figure 8e) and blood vessels formation (Figure 8f).

Figure 8 only represents the average values calculated from our semi-quantitative scoring. The complete scores obtained on the six animals can be seen in Table S2 in supporting information. The presence of osteoblasts is significantly higher in SiHA/VEGF scaffolds respect to the other samples. On the contrary NanoSiHA and NanoSiHA/VEGF scaffolds evidenced a lower presence of bone forming cells within their macropores compared to SiHA. Regarding the presence of bone resorbing cells, NanoSiHA and NanoSiHA/VEGF exhibited a higher presence of osteoclasts compared to SiHA and SiHA/VEGF, respectively, pointing out that nanosized bioceramics stimulate the osteoclast recruitment. Regarding inflammation, the scaffolds made of nanocrystalline SiHA show a higher trend for eliciting inflammatory response. In this sense, although Figure 8e does not show statistically significant differences, nanocrystalline scaffolds were scored with higher averaged values for inflammatory components. In addition, inflammatory components were absent in three animals out of six in SiHA and SiHA/VEGF scaffolds, whereas only one sheep out of six did not develop inflammatory response in NanoSiHA and NanoSiHA/VEGF. Finally, scaffolds containing VEGF induced a higher vascularization degree respect to those scaffolds without the factor (Figure 8f).

## 4. Discussion

Silicon substituted hydroxyapatites have demonstrated enhanced bioactive behavior compared to non-substituted hydroxyapatites under *in vivo* conditions [11,51]. Several studies have been carried out on the properties of silicon substituted HA as a function of the silicon content and regarding the sintering temperature at which these compounds are treated during the final stage of their processing [10,11,16,17,31,52]. In this work, the scaffolds made of silicon substituted hydroxyapatite and treated at high temperature (SiHA scaffolds) have a chemical composition and microstructure very similar to that already evaluated by other authors [11,51]. Therefore, SiHA scaffolds have been considered as positive control in order to evaluate the effect of decreasing the crystallite size (NanoSiHA), functionalizing with VEGF (SiHA/VEGF scaffolds) and both modifications simultaneously (NanoSiHA/VEGF scaffolds).

The results of our experiments indicate that we can manufacture macroporous scaffolds made of SiHA by the robocasting method, which show almost identical morphology and macroporous architecture but having very different microstructures and crystallite sizes. XRD experiments evidence that both SiHA and NanoSiHA are made of a unique apatite-like phase. However, the coherent diffraction domain (commonly referred as crystallite size) of SiHA is three times larger than that in NanoSiHA and the differences in grain size observed by SEM are even larger. The potential interest of nanosized SiHA arises from the positive results demonstrated by other authors with different nanosized bioceramics [18-20, 22-27], as well as by the assumption that the higher surface area of nanosized bioceramics could facilitate the release of osteogenic species (such as calcium cations and soluble silica species) and also foster scaffold degradation. Our results demonstrate that NanoSiHA scaffolds exhibit higher surface area and porosity and are degraded in higher extent than SiHA after 12 weeks of implantation. On the other hand, our results also show that VEGF can be associated to the scaffolds in a stable manner by using a simple solution and impregnation method. Both SiHA and NanoSiHA required no more than 10 min to adsorb almost 100% of the solved VEGF in a non-covalent way.

*In vitro* cell culture tests evidence that microstructure and functionalization with VEGF influence endothelial and pre-osteoblast cells in a different manner. Specifically, $EC_2$ endothelial cells proliferation is not affected by the grain size of the bioceramics. $EC_2$ cells attach, spread and proliferate similarly on both SiHA and NanoSiHA scaffolds. However, $EC_2$ proliferation is fostered by the presence of VEGF on the scaffolds surface, evidencing the biological activity of this factor and indicating that the non-covalent immobilization procedure did not alter the activity of VEGF as previously observed by other authors [53-55]

On the contrary, MC3T3-E1 pre-osteoblasts are very sensitive to the microstructural parameters. Pre-osteoblasts cannot proliferate on the scaffolds made of nanocrystalline SiHA despite of the similar initial attachment and spreading observed by SEM. Previous studies demonstrated the non-cytotoxicity of NanoSiHA [24], suggesting that the poor proliferative behavior of osteoblasts could be due to anoikis, i.e. apoptosis driven by cell detachment from weakly consolidated surfaces as that presented by NanoSiHA. The presence of VEGF favors the biological activity of pre-osteoblasts by enhancing the differentiation process, although this effect only occurs on highly crystalline SiHA/VEGF. In this sense, our results agree with other authors who reported on the increase of ALP activity and highlighted that VEGF promotes the differentiation of osteoblasts [37,39], thus opening the possibility of using VEGF to promote a healing in bone defects, but only in those cases where previous pre-osteoblast proliferation can occur.

Concerning the release kinetics of the VEGF from these scaffolds, *in vitro* test demonstrates that, after 96 hours, VEGF is still retained almost completely, since less than 0.02% could be detected by ELISA, in agreement with previous studies carried out on silicon substituted hydroxyapatites [42]. Previous works involving VEGF adsorbed in calcium phosphate scaffolds have evidenced the strong affinity of VEGF by these bioceramics [40]. The lack of release in our experiment would indicate that VEGF remains active immobilized on the implants surface as proposed by Hu *et al* [55].

*In vivo* studies in an osteoporotic sheep model strongly correlate with the *in vitro* results. The functionalization of crystalline SiHA scaffolds with VEGF leads to a significant improvement in the osteogenic properties of these bioceramics. As evidenced *in vitro*, the presence of VEGF must stimulate endothelial cell proliferation and pre-osteoblast differentiation, thus resulting in enhanced vascularization and new bone formation observed *in vivo* with thicker trabeculae and higher presence of osteoblasts. Silicon substituted hydroxyapatites have previously demonstrated a better bioactive behavior than non-substituted hydroxyapatites. The use of 3D printing for preparing macroporous scaffolds together with the functionalization with VEGF seem to be a significant advance for bone regeneration purposes even under an adverse scenario such as osteoporosis. Our experiments indicate that local administration of VEGF from macroporous SiHA scaffolds couples vascularization with new bone formation. However, the presence of silicon within the apatite structure and the immobilization of VEGF on the scaffolds surface is not enough to guarantee bone defect healing. Pre-osteoblasts require consolidated surfaces to proliferate. Certainly, those scaffolds made of nanocrystalline SiHA degrade faster and should be a source of $Ca^{2+}$, phosphate and silicon species at the defect local environment. Although the ionic release and scaffold degradation could be desirable for total bone regeneration, pre-osteoblasts cannot proliferate on these nanocrystalline and poorly consolidate surfaces. Therefore, fibrous tissue colonizes the scaffolds avoiding bone ingrowth. Moreover, particle release seems to promote osteoclast recruitment and an increase in the number of animals showing infiltration of inflammatory components. This *in vivo* response discards the use of NanoSiHA for bone regeneration purposes.

## 5. Conclusions

Macroporous scaffolds made of SiHA have been prepared with two different microstructures: nanocrystalline and crystalline. These scaffolds can be easily functionalized with VEGF by means of a simple impregnation method.

The presence of this factor on the scaffolds surface enhances the proliferation of endothelial cells, whereas highly crystalline microstructure favors pre-osteoblast proliferation and differentiation under *in vitro* conditions.

Scaffolds prepared with nanocrystalline SiHA evidence a deleterious response for the treatment of bone defects *in vivo*. Nanocrystalline microstructure leads to decreased bone ingrowth, thinner trabeculae, less presence of osteoblasts and higher presence of osteoclasts.

On the contrary, the association of VEGF with scaffolds made of highly crystalline SiHA exhibits better results in terms of volume of newly formed bone, trabeculae thickness and implant vascularization. These results evidence that the association of VEGF with macroporous scaffolds made of highly crystalline SiHA is a very promising alternative for the treatment of bone defects under osteoporosis conditions.

**Acknowledgements**

This study was supported by research grants from the Ministerio de Economía y Competitividad (project MAT2016-75611-R AEI/ FEDER, UE). M.V.-R. acknowledges funding from the European Research Council (Advanced Grant VERDI; ERC-2015-AdG Proposal 694160.